\documentstyle[11pt,paspconf]{article}

\begin{document}

\title{Dust temperature and the 
submillimeter--radio flux density ratio as a redshift indicator} 

\author{A. W. Blain}
\affil{Cavendish Laboratory, Madingley Road, Cambridge, CB3 0HE, UK} 




\begin{abstract}
It is difficult to identify the distant galaxies selected in existing submillimeter 
(submm)-wave surveys, because their positional accuracy is only several 
arcseconds. Currently, centimeter-wave 
VLA observations are required in order to 
determine sub-arcsec positions, and so to make reliable optical identifications. 
Carilli \& Yun (1999) pointed out that the ratio of the radio and submm-wave 
flux densities provides a redshift indicator for dusty star-forming galaxies, 
when compared with the tight correlation observed between the far-infrared 
(FIR) and radio flux densities for low-redshift galaxies. This method provides 
a useful, albeit imprecise, indication of the distance to a submm-selected 
galaxy. However, because the degeneracy between the effects of increasing the 
redshift of a galaxy and decreasing its dust temperature is not broken, it does 
not provide an unequivocal redshift estimate.
\end{abstract}




\section{Introduction}

The intensity of synchrotron radio emission from shock-heated electrons in
star-forming galaxies is known to be correlated tightly with their FIR
emission from dust grains heated by the interstellar radiation field
(see the review by Condon 1992). This FIR--radio correlation arises 
because both radiation processes are connected with the rate of 
ongoing high-mass 
star formation activity in a galaxy. The correlation links the flux densities 
of a galaxy in both the 60- and 100-$\mu$m {\it IRAS} passbands and at a 
frequency of 1.4\,GHz in the radio waveband. A reasonable pair of template 
spectral energy distributions (SEDs), which describe dusty galaxies at the 
relevant frequencies are shown in Fig.\,1, and compared with  
observations of the luminous dusty 
galaxy Arp\,220. 
This SED, galaxy 
evolution models taken from the same paper and the FIR--radio correlation can 
be combined to predict the faint counts of radio galaxies. The predicted 
count of galaxies brighter than 10\,$\mu$Jy at 8.4\,GHz is 0.8\,arcmin$^{-2}$, in 
agreement with the observed value of $1.0 \pm 0.1$\,arcmin$^{-2}$ (Partridge 
et al. 1997).

\begin{figure}[t]
\begin{center}
\plotfiddle{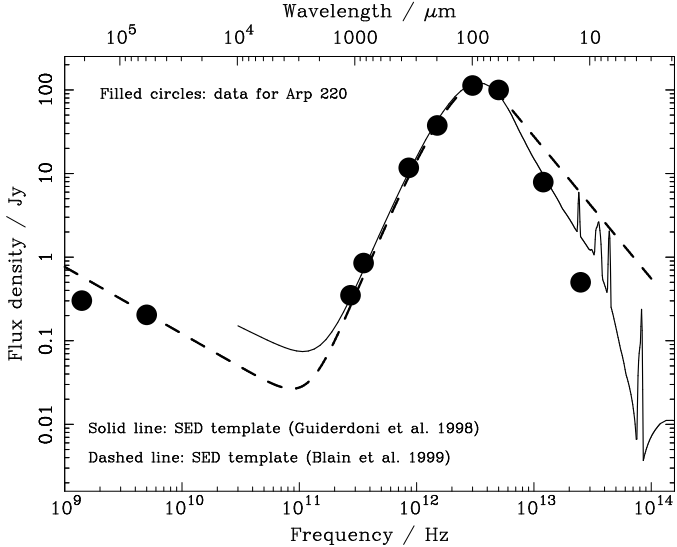}{6.1cm}{-90}{40}{40}{-150}{215}
\end{center}
\caption{The template SEDs used to describe dusty galaxies by 
Guiderdoni et al.\ (1998) and Blain et al.\ (1999b). The dashed line is 
normalized to match the FIR--radio correlation (Condon 1992). 
At long wavelengths, the SED is dominated by synchrotron radio 
emission with a spectral index $\alpha_{\rm radio} \simeq -0.8$. At short 
wavelengths the SED is dominated by the thermal radiation from dust 
grains with an emissivity index $\beta = 1.5$ and a temperature $T=38$\,K. 
The observed SED of Arp\,220 is also shown. Guiderdoni et al.'s 
template includes MIR spectral features (see Xu et al.\ 1998).
} 
\end{figure} 

The slope of the SED changes abruptly at a wavelength of about
3\,mm, at which the dominant contribution to the SED changes from
synchrotron emission to thermal dust radiation. Because the FIR--radio
correlation links the flux densities on either side of this spectral break, it
could be exploited to indicate the redshift of the galaxy (Carilli \& Yun 1999).
Carilli \& Yun calculated that the radio--submm flux density ratio of a distant 
dusty galaxy, which lies on the FIR--radio correlation, should be
\begin{equation}
{ {S_{\rm 1.4GHz}} \over { S_{\rm 850{\mu}m} } } = 3.763
(1+z)^{1.007(\alpha_{\rm radio} - \alpha_{\rm submm})},
\end{equation}
as a function of redshift $z$. $\alpha_{\rm radio}$ and 
$\alpha_{\rm submm}$ are the spectral indices of the SED, 
$S_\nu \propto \nu^\alpha$, in the radio and submm wavebands. 
Typically, $\alpha_{\rm radio} \simeq -0.8$ and $\alpha_{\rm submm} \simeq 3.0$ 
to 3.5. $\alpha_{\rm submm}$ is the sum of the Rayleigh-Jeans spectral index 
($\alpha=2$) and $\beta$, the spectral index in the dust emissivity function 
$\epsilon_\nu \propto \nu^\beta$. Carilli \& Yun (1999) also derived empirical flux 
density ratio--redshift relations from the SEDs of Arp\,220 and M82. The
spread of the redshift values that correspond to a fixed flux density ratio
across their four models corresponds to an uncertainty $\Delta z \simeq 0.5$.

The sensitive 850-$\mu$m SCUBA camera at the JCMT (Holland et al.\ 1999) 
has been used to detect high-redshift dusty galaxies (see Blain et al.\ 1999a 
and references within). These are probably the high-redshift counterparts to the 
low-redshift ultraluminous infrared galaxies (ULIRGs), and so their SEDs 
might also be expected to follow the FIR--radio correlation. In this case, the 
ratios of their radio and submm-wave flux densities could be substituted 
into equation (1) to indicate their redshifts.

\begin{table}
\caption{Spectroscopic ($z_{\rm spec}$) and radio-submm redshifts 
for dusty galaxies calculated 
using the formula of Carilli \& Yun (1999). The 
predicted redshifs $z_{3.0}$ and $z_{3.5}$ are presented assuming 
$\alpha_{\rm submm}$ values of 
3.0 and 3.5 respectively.} 
\begin{center}\scriptsize
\begin{tabular}{ccccc}
Name & $z_{\rm spec}$ & $z_{3.0}$ & $z_{3.5}$ & Reference\\ 
\tableline
SMM\,J02399$-$0136 &2.80& 
$2.9 \pm 0.3$ & $2.3 \pm 0.3$ & Frayer et al.\ (1998)\\
SMM\,J14011+0252 & 2.57&  
$3.8 \pm 0.4$ & $3.3 \pm 0.4$ & Frayer et al.\ (1999)\\
BR\,1202$-$0725 & 4.69 & 
$3.9 \pm 0.5$ & $3.1 \pm 0.4$ & Kawabe et al.\ (1999)\\
{\it IRAS} F10214+4724 & 2.29 & 
$2.8 \pm 0.2$ & $2.2 \pm 0.3$ & Rowan-Robinson et al.\ (1993)\\
Arp\,220 & 0.02 & $1.1 \pm 0.1$ & $1.0 \pm 0.1$ & ...\\ 
\end{tabular}
\end{center}
\end{table} 

The observed radio and submm-wave flux densities of five dusty 
galaxies are used to predict their redshifts in Table\,1.  
This technique provides a coarse indication of the redshift
of these five galaxies, which all lie at spectroscopic redshifts
within the bounds of the range of predictions made by Carilli \& Yun's
four models.
What are the systematic effects that limit the reliability
of the inferred redshifts?

\section{The FIR--radio correlation at high redshifts}

The FIR--radio correlation is based on observations of a range of 
low-redshift galaxies; low- and high-luminosity spiral galaxies, irregular 
star-forming dwarf galaxies and ULIRGs. One factor that could modify the 
general properties of the SEDs of high-redshift dusty galaxies is the $(1+z)$
increase in the temperature of the cosmic microwave background  
(CMB) from its value at $z=0$. However, unless the dust 
temperature $T \le 20$\,K this is 
unlikely to have a significant effect on the SED if $z \le 5$. The details 
are discussed in more detail by Blain (1999a). 

The increasing temperature of the CMB has another effect. Above a certain
redshift the energy density in the CMB exceeds that
in the magnetic field in the interstellar medium (ISM) of the observed 
galaxy. Synchrotron radio emission will thus be suppressed due to the 
cooling of relativistic electrons by inverse Compton scattering of CMB 
photons. Carilli \& Yun (1999) estimate that this will typically occur for 
ULIRGs at $z \ge 6$, and at $z \ge 3$ for the lesser magnetic fields 
in the ISM of the Milky Way. This effect is not included in the results 
presented here; however, this potential deficit in the
radio flux density from high-redshift galaxies should be borne in mind.

\section{Uncertainties}

\begin{figure}[t] 
\begin{center}
\plotfiddle{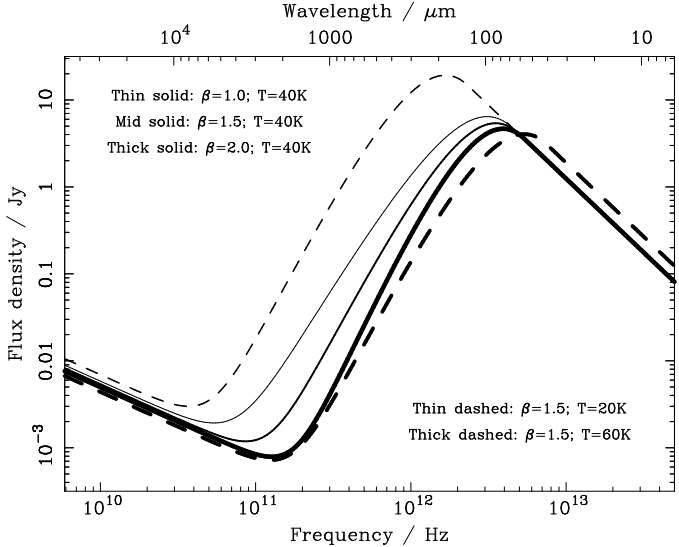}{6.1cm}{-90}{40}{40}{-150}{215}
\end{center}
\caption{Examples of SEDs defined by different values of the $z=0$ dust
temperature $T$ and dust emissivity index $\beta$, all normalised to the same
flux density at 60\,$\mu$m. The SEDs are very different in the submm
waveband, but very similar in both the FIR and radio wavebands.
}
\end{figure}

Galaxies with a range of different dust temperatures $T$ 
lie on the FIR--radio correlation. This is because the FIR flux densities 
involved in the correlation are at wavelengths close to the peak of the SED of 
a typical dusty galaxy. Hence, shifting the position of the peak of the SED,
by changing $T$, makes little difference to the integrated 
FIR luminosity and thus to the FIR--radio flux density
ratio. However, in the submm waveband, the effect of modifying either the
dust temperature $T$ or the emissivity index $\beta$ is much greater, as
shown by the five different model SEDs in Fig.\,2. Three different dust
temperatures $T=20$, 40 and 60\,K are included, each with an emissivity index
$\beta = 1.5$. In addition, the SED is calculated for a $T=40$\,K model with
$\beta = 1.0$ and 2.0.

\begin{figure}[t]
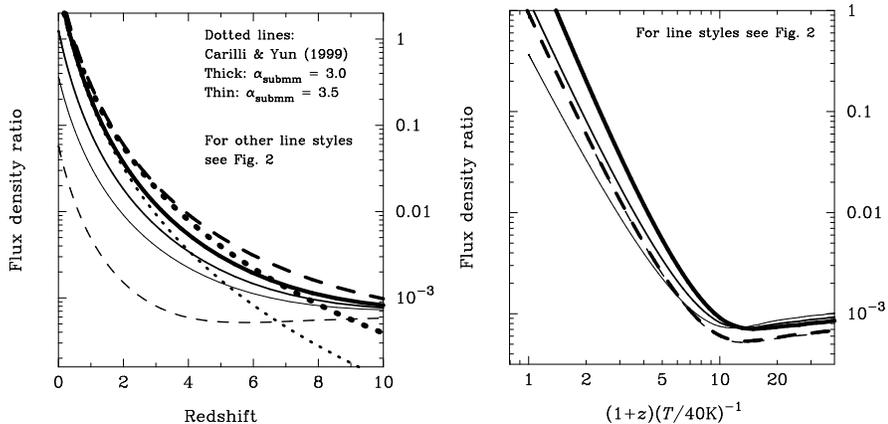
 
\begin{center}
\plotfiddle{blainaposter3a.ps}{4.3cm}{-90}{36}{36}{-230}{180}
\end{center}
\begin{center} 
\plotfiddle{blainaposter3b.ps}{4.3cm}{-90}{36}{36}{-60}{328}
\end{center} 
\vskip -5cm 
\caption{ 
Left (a): the 1.4-GHz:850-$\mu$m flux density ratios expected for the five
model SEDs shown in Fig.\,2 as a function of redshift $z$. A considerable
range of redshifts could be deduced from a measurement of this ratio if the
temperature $T$ or emissivity index $\beta$ of the dust in the galaxy being
observed was uncertain. The curves from Carilli \& Yun (1999) are plotted
assuming $\alpha_{\rm radio} = -0.8$. Carilli \& Yun stressed that a ratio 
greater than 0.1 is likely to indicate $z < 1$, while a ratio less than $10^{-2}$ 
is likely to indicate $z > 2$. Right (b): the same ratio plotted as a 
function of $(1+z)/T$, showing that it is difficult to break the degeneracy 
between $z$ and $T$ using joint radio--submm measurements.
}
\end{figure}

The predicted forms of the submm--radio flux density ratio as a function 
of redshift for all five template SEDs are shown in Fig.\,3(a). The results of 
Carilli \& Yun (1999), calculated using equation (1), are also shown for 
comparison. Carilli \& Yun's equation provides a good description of the 
1.4-GHz:850-$\mu$m flux density ratio if $T \ge 60$\,K. 
However, because the positions of the curves in Fig.\,3(a) are quite different 
if $T \le 60$\,K, the redshift that would be assigned to a 
cooler galaxy using the 1.4-GHz:850-$\mu$m flux density ratio alone is 
uncertain. This is true even if the FIR--radio correlation is assumed to be free 
from any intrinsic scatter, and if any additional contribution to the radio flux 
density from an active galactic nucleus (AGN) is neglected. The intrinsic 
scatter in the FIR--radio correlation is about 0.2\,dex, and so because the 
power-law index of the lines in Fig.\,3(a) is about $-2$ at $z \sim 2$, an 
additional 0.1\,dex ($\simeq 25$\,per cent) uncertainty would be expected. See 
Carilli \& Yun (1999) for a discussion of the effects of any AGN component, 
which should lead to a conservatively low redshift estimate.

It is interesting to replot the curves in Fig.\,3(a) as a function of the combined 
redshift--dust temperature parameter $(1+z)/T$. This is the quantity that is
constrained by measuring the position of the peak of the thermal
dust emission component of the SED by combining observations in the
FIR and mid-infrared (MIR) wavebands. The results are shown in
Fig.\,3(b). Because the radio flux density is produced by a non-thermal
emission mechanism, a measurement of the radio--submm flux density
ratio might be expected to break the degeneracy between temperature and
redshift. However, because the differences between the curves in Fig.\,3(b) are
not much greater than the scatter in the FIR--radio correlation, the
degeneracy remains.

\section{Other redshift indicators}

The `redshifted dust temperature' of a galaxy, $T/(1+z)$, can be determined by
measuring the frequency of the peak of the FIR dust component of
the SED (see Figs\,1 and 2), but $z$ and $T$ 
cannot be determined independently. Locating the peak frequency requires 
both long and short wavelength observations: see Blain (1999b) and Hines 
(1999) elsewhere in this volume. 

Spectral features produced by emission from polycyclic aromatic
hydrocarbon molecules and atomic fine-structure lines in the restframe
MIR waveband 
(see Fig.\,1) could be exploited to obtain photometric redshifts for distant
galaxies (Xu et al.\ 1998). At shorter wavelengths, the prospects for
obtaining photometric redshifts using the 3--10\,$\mu$m {\it SIRTF} IRAC
camera have been discussed recently by Simpson \&
Eisenhardt (1999). Photometric redshifts deduced from these features will not 
be subject to the $T$--$z$ degeneracy.

The identification of an optical counterpart to a submm-selected
galaxy usually requires a radio observation and a great deal of observing
time (see for example Ivison et al.\ 1998). The optical magnitudes and
colours of heavily-obscured submm-luminous galaxies are expected to
extend over wide ranges, and so the derivation of redshift information
from broad-band optical photometry will probably require careful individual 
analysis of each submm-selected galaxy. 
Photometry and spectroscopy of likely optical counterparts to
submm-selected galaxies (Smail et al.\ 1998; Barger et al.\ 1999)
indicate that the ratios of 850-$\mu$m and optical $I$-band flux densities
are scattered by about an order of magnitude across the sample.

\section{Conclusions}

Radio observations of submm-selected galaxies are crucial in order to
make reliable optical identifications. The ratio of the radio and
mm/submm-wave flux densities also provides information about the 
redshift of the galaxy (Carilli \& Yun 1999). However, the temperature 
and emissivity of dust in distant galaxies has a very significant effect on 
the ratio of the submm-wave and radio flux densities, and so for 
reasonable values of the dust emissivity and temperature, this ratio 
cannot be used to break the degeneracy between the dust temperature 
and redshift of a distant dusty galaxy.

\acknowledgements

I thank Chris Carilli, Kate Isaak, Rob Ivison, Richard McMahon, Kate Quirk
and Min Yun for helpful comments, and OCIW for support at this meeting. 


%
%

%

\end{document}